\documentclass[12pt,a4paper]{article}
\usepackage{type1cm,amsmath,hangcaption,graphicx,indentfirst}
\usepackage[psamsfonts]{amssymb}
\numberwithin{equation}{section}

\begin{document}
\begin{titlepage}

 \renewcommand{\thefootnote}{\fnsymbol{footnote}}
\begin{flushright}
 \begin{tabular}{l}
 KEK-TH-1322\\
 IPMU-09-0083\\
 arXiv:0907.3832\\
 \end{tabular}
\end{flushright}

 \vfill
 \begin{center}

 \vskip 2.5 truecm

\noindent{\large \textbf{Topological String on OSP(1$|$2)/U(1)}}\\
\vspace{1.5cm}

\noindent{ Gaston Giribet$^a$\footnote{E-mail: gaston@df.uba.ar}, Yasuaki Hikida$^b$\footnote{E-mail:
hikida@post.kek.jp}
and Tadashi Takayanagi$^c$\footnote{E-mail: tadashi.takayanagi@ipmu.jp}}
\bigskip

 \vskip .6 truecm
\centerline{\it $^a$Physics Department, University of Buenos Aires, and Conicet,}
\centerline{\it Ciudad Universitaria, Pab. I, 1428, Buenos Aires, Argentina}
\bigskip
\centerline{\it $^b$KEK Theory Group, Tsukuba, Ibaraki 305-0801, Japan}
\bigskip
\centerline{\it $^c$Institute for the Physics and Mathematics of the
Universe (IPMU),} \centerline{\it University of Tokyo, Kashiwa,
Chiba 277-8582, Japan}

 \vskip .4 truecm

 \end{center}

 \vfill
\vskip 0.5 truecm

\begin{abstract}

We propose an equivalence between topological string on
OSP(1$|$2)/U(1) and $\hat{c}\leq 1$ superstring with ${\cal N}=1$
world-sheet supersymmetry. We examine this by employing a free field
representation of OSP(1$|$2) WZNW model and find an agreement on the
spectrum. We also analyze this proposal at the level of scattering
amplitudes by applying a map between correlation functions of
OSP(1$|$2) WZNW model and those of ${\cal N}=1$ Liouville theory.

\end{abstract}
\vfill
\vskip 0.5 truecm

\setcounter{footnote}{0}
\renewcommand{\thefootnote}{\arabic{footnote}}
\end{titlepage}

\newpage

\tableofcontents
\section{Introduction}
\label{Intoduction}

Superstrings on AdS spaces have been widely studied by virtue of their
applications to the holographic duality, i.e. the AdS/CFT correspondence
\cite{Ma}; and it has become clear that the structure of supergroup
$\sigma $-models is of great importance to investigate superstrings
in these spaces. For instance, the supergroup
PSU(2,2$|$4) turns out to be important to construct superstring theory
on $AdS_5 \times S^5$ \cite{MT}. Besides, superstring theory on $AdS_3 \times
S^3$ can be described in terms of the PSL(1,1$|$2) WZNW model \cite{BVW}.
However, in spite of its importance, quantizing supergroup $\sigma $-models
is a quite difficult problem, and hence solving superstring theory
on AdS spaces exactly still remains as an unsolved question.

Fortunately, there is a simpler type of duality for which string
world-sheet theory is still described by a supergroup WZNW model. It
has been established in \cite{TT,DKKMMS} that two-dimensional
superstring (type 0 string) can be holographically described by a
simple Hermitian matrix model. At present, this is the only
dynamical model of string theory which is non-perturbatively
well-defined and is exactly solvable even at finite temperature. The
two-dimensional type 0 string theory is originally defined by string
world-sheet theory with the $\hat{c}=1$ matter coupled to ${\cal N}
= 1$ super Liouville theory. Boosting the linear dilaton with
Liouville potential kept the same, this theory can be extended to
$\hat{c}\le 1$ type 0 string theory as it has been done for bosonic
string in \cite{Ta,Takayanagi}. Note that dual matrix model can be
constructed even for $\hat{c} < 1$ case, as shown in \cite{Ta}.

In this paper, we argue that these $\hat{c}\leq 1$  superstring
theories can be described by utilizing the supergroup
OSP(1$|$2).\footnote{Current superalgebra of OSP type
also appears in an attempt
\cite{OT} to generalize heterotic string so as to be dual to
Type I string theory with a OSP gauge symmetry.}
 Precisely speaking, we propose that the
$\hat c \leq 1$ superstring is equivalent to topological string
on ${\cal N}=2$ superconformal coset OSP(1$|$2)/U(1).%
\footnote{Topological strings on cosets based on sugerpgroups
have been studied for the analysis of Maldacena conjecture
via world-sheet theory in \cite{Berkovits1,Berkovits2,Bonelli1,Berkovits3,Bonelli2}.}
This
relation can be thought of as a supersymmetric version of the known
relation between $c \leq 1$ bosonic string theory and topological
string on SL(2)/U(1) \cite{MV,Takayanagi}. This extension might
be guessed from the quantum Hamiltonian reduction since
OSP(1$|$2) WZNW model is reduced to ${\cal N} = 1$ super Liouville
theory \cite{BO}, just like SL(2) WZNW model is reduced to bosonic Liouville theory \cite{BO2}.

This paper is organized as follows.
In the next section, we explicitly
construct the ${\cal N}=2$ superconformal coset OSP(1$|$2)/U(1)
as a natural extension of Kazama-Suzuki model for bosonic cosets
\cite{KS,KS2}.%
\footnote{Generic construction of Kazama-Suzuki model for 
cosets of supergoups was given in \cite{CRS,Creutzig} very recently.}
We analyze it in
the free field theory and show that the $\hat{c}\leq 1$ string
world-sheet appears after the topological twisting.
In particular, we show that free fields in the coset model
become the matter contents of $\hat{c} \leq 1$ superstring,
and the chiral primaries of the coset model are identified with
the physical operators of $\hat{c} \leq 1$ superstring.
In section 3,
we review and extend the map between the correlation
functions in OSP(1$|$2) WZNW model and those in the ${\cal N}=1$
Liouville theory.
This relation was originally obtained in \cite{HS2} as an
generalization of $H_3^+$-Liouville relation \cite{RT,HS3}.
In section 4, after briefly reviewing
$\hat{c}\le 1$ superstrings, we apply this map to study the
scattering S-matrices. We explicitly show that the
correlation functions of physical operators in the topological
model are mapped to those of physical operators in the
$\hat{c} \leq 1$ superstring.
In section 5, we summarize the conclusion.
In the appendix, we discuss correlation
functions of  OSP(1$|2$) WZNW model in the free field
representation.

\section{${\cal N}=2$ Coset OSP(1$|$2)/U(1) and 2D Superstring}
\label{coset}

In this section we construct and analyze ${\cal N}=2$
supersymmetric coset
(Kazama-Suzuki model \cite{KS,KS2}) based on OSP(1$|$2)/U(1). After its
topological twisting, we show explicitly from the free field
theory analysis that the world-sheet theory of $\hat{c}\leq 1$
superstring indeed appears. We also discuss chiral primary
states which are the physical states in the topologically twisted
theory.

\subsection{OSP(1$|$2) Current Algebra}

The current algebra of OSP(1$|$2) includes SL(2) bosonic subalgebra,
which is generated by $J^{3} (z)$ and $J^{\pm} (0)$ with their OPEs%
\footnote{For a while we concentrate on the holomorphic part.}
\begin{align}
 J^+(z) J^-(0) \sim  \frac{k}{z^2} - \frac{2J^3(0)}{z} , \qquad
 J^3(z) J^{\pm}(0) \sim \pm \frac{J^\pm (0)}{z} ,\qquad
 J^3(z)J^3(0)\sim -  \frac{k}{2z^2} .
\end{align}
In addition to these bosonic generators, there are fermionic ones with
\begin{align}
 &J^3(z) j^{\pm}(0)\sim \pm\frac{j^\pm (0)}{2z} , \qquad
 J^{\pm} (z) j^\mp (0) \sim  \mp \frac{j^\pm(0)}{z} , \\  \nonumber
 &j^+(z)j^-(0) \sim  \frac{2k}{z^2} - \frac{2J^3(0)}{z} , \qquad
 j^{\pm}(z) j^{\pm}(0) \sim - \frac{2J^{\pm}(0)}{z} .
\end{align}
The energy momentum tensor is given by Sugawara construction and the
central charge is $c = 2k/(2k-3)$. These are the definition of
 OSP(1$|$2) current algebra with level $k$.

In a free field representation \cite{Wakimoto,BO,ERS,ERS2} the above
currents may be expressed as
\begin{align}
\label{freefield}
 &J^- = \beta , \qquad J^+ = \beta \gamma ^2 - \frac{1}{b}
 \gamma \partial \phi + \gamma \theta p + k \partial \gamma
  - (k -1 ) \theta \partial \theta  , \\
 &J^3 = \beta \gamma - \frac{1}{2b} \partial \phi + \frac12 \theta p  , \qquad
 j^- = p - \beta \theta , \qquad j^+ =
 \gamma p - \beta \gamma \theta
+ \frac{1}{b} \theta \partial \phi - (2k -1) \partial \theta ,  \nonumber
\end{align}
where the OPEs of these free fields are
\begin{equation}
 \phi (z) \phi (0) \sim - \ln z , \qquad
 \beta (z) \gamma (0) \sim \frac{1}{z} , \qquad
 p(z) \theta (0) \sim \frac{1}{z} ~.  \label{Propagators}
\end{equation}
The field $\phi$ has the background charge $Q_\phi =b$
and the central charge is $c = 1 + 3 Q_{\phi }^2$.
Here the parameter $b$ is related to the level $k$ as $1/b^2 = 2k - 3$.
The bosonic fields $(\beta,\gamma)$ have conformal weights
$(1,0)$ and the central charge of this system is $c=2$.
On the other hand, $(p,\theta)$ are fermions with conformal weights
$(1,0)$ and central charge $c = -2$.
In the following analysis, it is useful to bosonize the fermionic fields $(p,\theta)$ as
\begin{equation}
\theta = e^{i Y} , \qquad p = e^{ - i Y} . \label{Bosonization}
\end{equation}
For instance, the $J^3$ current takes the form
\begin{equation}
J^3  = \beta \partial \gamma -\frac{1}{2b} \partial \phi +\frac{i}{2}
\partial Y .\label{J3Y}
\end{equation}
The energy momentum tensor is given by
\begin{align}
 T = \beta \partial \gamma - \frac12 \partial \phi \partial \phi
 + \frac{b}{2} \partial^2 \phi - p \partial \theta
  = \beta \partial \gamma - \frac12 \partial \phi \partial \phi
 + \frac{b}{2} \partial^2 \phi -  \frac12 \partial Y \partial Y
 + \frac{i}{2} \partial^2 Y  \label{TheT}
\end{align}
in terms of these free fields.

One of the merits to utilize the free field representation is
that vertex operators can be expressed in a simple form.
Namely, the vertex operators of OSP(1$|$2) model can be
written in terms of free fields as
\begin{align}
 \Phi^s_{j,m}  \sim e^{i s Y} \gamma^{ j - s/2 +m }
 e^{ - 2 b j \phi} ,
 \label{freev}
\end{align}
whose conformal weight with respect to (\ref{TheT}) is
\begin{equation}
\Delta= - 2 b^2 j(j+\tfrac12) + \tfrac12 s (s-1) .  \label{ConformalDimension}
\end{equation}
Here we set $s=0,1/2,1$. For $s=0,1$ we can express the vertex operator
even in terms of $\theta$, but not for $s=1/2$. The operator with $s=1/2$
corresponds to a twist operator in R-sector, whose role was argued
for GL(1$|$1) WZNW model in \cite{LeClair}, see also \cite{PT}.
In order to compute correlation functions, the overall normalization
of vertex operators should
be fixed. More precise definition will be given in section \ref{ribault}.
Correlation functions of vertices (\ref{freev}) are discussed in the appendix, where the Coulomb gas prescription is given.

Notice that the expression (\ref{ConformalDimension}) is invariant under the Weyl
transformation $j\to -j-1/2$ and under $s\to 1-s$. This
allows us to consider a second contribution to (\ref{freev}) which
goes like $\sim e^{2b(j+1/2)\phi }$ and it would dominate the large $\phi $
regime for $j>-1/4$.
In addition, there exist conjugate representations which are similar to
those that exist in the free field realization of SU(2) model \cite%
{Dotsenko, Dotsenko2, ERS, ERS2}. For instance, one finds the operator
\begin{align}
\hat{\Phi }_{j,j+\frac{1}{2}}^{0}\sim \beta ^{k-2-2j}e^{b(2k-3-2j)\phi }
\end{align}
which represents a Kac-Moody primary of conformal dimension (\ref%
{ConformalDimension}), with $m=j+1/2$ and $s=0$. It can be also thought of
as a conjugate representation for the state with $m=j$, $s=1$.

In order to define the  OSP(1$|$2)/U(1) coset theory,
we utilize the
representation introduced in \cite{BK} and \cite{DVV} to realize the
$SL(2)/U(1)$ coset theory.
This amount to introduce
a boson $X^3(z)$ with $X^3 (z) X^3 (0) \sim - \ln z$,
as well as a $(b,c)$ ghost system, which are
used to mode out the $U(1)$ factor.
Then the vertex operators of the coset theory are given by
\begin{align}
\Phi^s_{j,m} =
 \Psi^s_{j,m}    e^{  - i \sqrt{\frac{2}{k}} m X^3 }  .
 \label{parafermion}
\end{align}
As we will discuss below, the OSP(1$|$2) current algebra admits
the symmetry under the spectral flow action as in the case of
SL(2) WZNW model \cite{MO}.
For  OSP(1$|$2) model, spectrally flowed states are defined
in this form as
\begin{align}
\Phi^{s,w}_{j,m} =
 \Psi^s_{j,m} e^{ - i \sqrt{\frac{2}{k}} (m + \frac{k}{2}w) X^3 }
 \label{spectralflow}
\end{align}
with the index of spectral flow $w$.

\subsection{${\cal N} = 2$ Supersymmetric Coset Model}

In this subsection we would like to construct
 ${\cal N} = 2$ supersymmetric model based on
the coset OSP(1$|$2)/U(1). For the purpose we first generalize
the OSP(1$|$2) current algebra into ${\cal N} = 1$ supersymmetric
version, therefore we need superpartners of currents
$(J^3,J^{\pm},j^{\pm})$. While we introduce fermions
$(\psi^3,\psi^{\pm})$ with spin $1/2$ for the bosonic currents
$(J^3,J^{\pm})$, we include bosons $\varphi^{\pm}$ with spin $1/2$
for the fermionic currents $j^{\pm}$. We assume the OPEs of these
fields as
\begin{align}
 \psi^+ (z) \psi^- (0) \sim \frac{1}{z} , \qquad
 \psi^3 (z) \psi^3 (0) \sim \frac{1}{z} , \qquad
 \varphi^+ (z) \varphi^- (0) \sim - \frac{1}{z} .
\end{align}
Notice that the extra bosons  $\varphi^{\pm}$ satisfy
wrong spin-statistic relation.
With these new fields we can define ${\cal N}=1$ supercurrents as
\begin{align}
 \hat{J}^\pm  = J^\pm + \sqrt{2} \psi^\pm  \psi^3 +
  \frac{1}{2}\varphi^\pm \varphi^\pm , \qquad
 \hat{j}^\pm = j^{\pm} \pm i \sqrt{2} \psi^\pm \varphi^\mp
  \pm i \psi^3\varphi^\pm ,
 \label{ospsuper}
\end{align}
along with
\begin{align}
 \hat{J}^3 =J^3+\psi^+\psi^-  + \frac{1}{2}\varphi^+ \varphi ^- .
\end{align}
We construct the coset model by using the last current as well as removing
one of the fermions $\psi^3$ to preserve ${\cal N}=1$ world-sheet supersymmetry.

We can show that the coset model actually has enhanced ${\cal N}=2$ supersymmetry as Kazama-Suzuki models for bosonic cosets
\cite{KS,KS2}.
We find that the generators of ${\cal N}=2$ superconformal
symmetry are
\begin{align}
\label{supercon}
&J_R= - \frac{1}{2k - 3}\left(2J^3+(2k-1)\psi^+\psi^- + (2k-2)\varphi^+\varphi^-\right) ,
\\
&G^{\pm} = \frac{1}{\sqrt{2k-3}} \left( 2 J^\pm \psi^\mp
 \pm \sqrt{2} j^\pm \varphi^\mp
 + (\varphi^\mp)^2 \psi^\pm \right) , \nonumber
\\
&T=  \frac{1}{2k-3}\Bigl[J^+J^-+J^-J^++\frac{1}{2}(j^-j^+-j^+j^-)
 + 4J^3\psi^+\psi^- + 2J^3\varphi^+\varphi^-
+ 2\psi^+\psi^-\varphi^+\varphi^- \nonumber \\
& \qquad - \frac{2k+1}{2}(\psi^+\partial\psi^-+\psi^-\partial\psi^+)
 - (k-1)(\varphi^+\partial\varphi^- - \varphi^-\partial
\varphi^+) + \frac{1}{2}(\varphi^+)^2(\varphi^-)^2\Bigr] . \nonumber
\end{align}
In fact, we can compute the OPEs of generators as
\begin{align}
\label{algentwo}
&
T(z)T(0)\sim \frac{c/2}{z^4}+\frac{2T(0)}{z^2}+\frac{\partial T(0)}{z} ~ , \qquad
T(z)G^\pm(0)\sim \frac{\frac{3}{2}G^\pm(0)}{z^2}
 +\frac{\partial G^\pm (0)}{z} , \\
&T(z)J_R(0)\sim \frac{J_R(0)}{z^2}+\frac{\partial J_R (0)}{z} , \qquad
 J_R(z)G^\pm (0)\sim \pm \frac{G^\pm (0)}{z} , \qquad
 J_R(z)J_R(0)\sim \frac{c/3}{z^2} , \nonumber \\
 & G^\pm (z)G^\mp (0) \sim \frac{\frac{2}{3}c}{z^3} \pm \frac{2J_R  (0)}{z^2}
+\frac{2T (0)}{z}
\pm \frac{\partial J_R (0)}{z}  ,
\qquad G^\pm (z)G^\pm (0) \sim 0 .
\nonumber
\end{align}
In this way we have explicitly shown that
these generators satisfy the ${\cal N}=2$ superconformal
algebra with central charge $\hat{c}= c/3 = 1/(2k-3)$.

\subsection{Topological Twisting}

To realize the U(1) quotient more explicitly, we combine a free
scalar field $X$ with the ${\cal N}=1$ OSP current algebra and
finally take a quotient by the complexified U(1) (i.e. the complex
plane ${\mathbb C}$). This quotient can be done by taking BRST
invariant state about the BRST operator
\begin{align}
\label{brst}
Q_{B}=\int dz C(z)J_{g}(z) ,
\end{align}
where we
introduced fermionic ghosts $(B,C)$ with
the conformal weights $(1,0)$.
Here the BRST current $J_{g}$ is defined by
\begin{align}
\label{defghi}
J_g =\hat{J}^3  - \frac{i}{2b}\partial X ,
\end{align}
and it is easy to see that
$J_g(z)J_g(0)\sim 0$ which guarantees $Q_{B}^2=0$.

Notice that in this formalism we can always set $J_g(z)$ to zero
since it is gauged.
Using this fact we can use the following expression of the
R-current of ${\cal N}=2$ superconformal algebra as
\begin{align}
\label{rtcurrent}
J'_R =J_R -\frac{4k - 8}{2k - 3}J_g
=-2J^3-3\psi^+\psi^- - 2\varphi^+\varphi^-
+ i \frac{2k- 4}{\sqrt{2k-3}}\partial X.
\end{align}
In the anti-holomorphic part, we use the same expression with bars.%
\footnote{
This choice means that we gauge the vector $U(1)$ current instead of the
axial $U(1)$ current. The former may produce a trumpet like
geometry and the latter a black hole like geometry \cite{DVV}.
We choose the vector gauge just for the simplicity of expression.}
We will find that this form of R-current is useful to construct
the topological model as in the bosonic case \cite{MV,Takayanagi}.

Now, we perform topological twists \cite{EY,WTST} by
using the expression \eqref{rtcurrent} of R-current.
Namely, we redefine the energy momentum tensor by
$T^{top}=T+\frac{1}{2}\partial J'_R$ and $\bar T^{top} = \bar T +
\frac{1}{2} \bar \partial \bar J '_R$.
Employing the free field representation \eqref{freefield},
we then find the following maps of fields.
First of all, the background charge of the field $\phi$ is
shifted from $Q_\phi = b$ to $Q_\phi = b+1/b$. After the
twist, the field $\phi$ corresponds to the Liouville field.
Recall that the central charge is
written as $c = 1 + 3 Q^2$ in terms of background charge $Q$.
Next, the field $X$ would have background charge
$Q_X = i (1/b - b)$ after the twist, and this field
becomes the bosonic part of the $\hat c \leq 1$ matter.
The conformal weights of fermions $(\theta, p)$ are shifted
from $(0,1)$ to $(1/2,1/2)$ and they become superpartners
of the above bosonic fields. The other fields $\psi^\pm$ and
$\varphi^\pm$ are mapped to the  superghosts
$(b,c)$ and $(\beta ' , \gamma ')$ of type 0 superstring theory.%
\footnote{Here we use the notation $(\beta ' , \gamma ')$
to represent superghosts of superstring in order to
distinguish them from the ones in \eqref{freefield}.}
In table \ref{ba} the changes of conformal weights are summarized.
In the end we expect that $(\beta,\gamma)$ would be canceled out with
$(B,C)$ as in the bosonic string case \cite{MV}. In this way
we obtain the same field contents as the world-sheet theory of
the type 0 $\hat{c} \leq 1$ string including ghosts.
More detailed explanation of the type 0 string
will be given in subsection \ref{secc}.
\begin{table}
\begin{center}
\begin{tabular}{|c|c|c|c|c|}
\hline
 ~ & \multicolumn{2}{c|}{Before twisting}
 & \multicolumn{2}{c|}{After twisting} \\
 \cline{2-5}
 ~ & Central charge& Conformal weights &
  Central charge & Conformal weights \\
\hline
$(\theta , p)$ & $-2$ & $(0,1)$ &
$1$ & $(1/2,1/2)$ \\
\hline $(\psi^+,\psi^-)$ & $1$ & $(1/2,1/2) $ &
$-26$ & $(2,-1)$ \\
\hline
$(\varphi^+,\varphi^-)$ &
$-1$ & $(1/2,1/2)$ &
$11$ & $(3/2,-1/2)$ \\
\hline
$(\beta,\gamma)$ &
$2$ & $(1,0)$ &
$2$ & $(0,1)$ \\
\hline
$(B,C)$ &
$-2$ & $(1,0)$ &
$-2$ & $(1,0)$ \\
\hline
\end{tabular}
\end{center}
\caption{Changes of central charges and conformal weights
after topological twisting.}
\label{ba}
\end{table}

\subsection{Chiral Primaries}

In the previous subsection we have shown that free fields
in the ${\cal N} = 2 $ coset are mapped to the matter contents
of the $\hat c \leq 1$ superstring theory after the topological twist.
In fact we can identify physical operators
of the topological model with those of the $\hat c \leq 1$ superstring,
which is the subject of this subsection.
In
order to define the coset model we have introduced two spin $1/2$ fermions
$\psi^{\pm}$ and two spin $1/2$ bosons $\varphi^\pm$. With the bosonization
formula, they can be written as
\begin{align}
 \psi^+ = e^{iH} , \qquad \psi^- = e^{-iH}  , \qquad
 \varphi^+ =  e^{-  \chi}  \partial \xi  ,  \qquad
 \varphi^- =  e^{ \chi} \eta  ,
 \label{bosonized}
\end{align}
where $H , \chi$ are free bosons without background charges
and free fermions $\xi,\eta$ are with
$\Delta_\xi = 0, \Delta_\eta = 1$. The non-trivial OPEs are
given as
\begin{align}
 H(z) H(0) \sim - \ln z , \qquad
 \chi (z) \chi (0) \sim - \ln z , \qquad
  \eta (z)  \xi (0) \sim \frac{1}{z} .
\end{align}
These bosonized expressions of fermions are useful to
define vertex operators.
Since the operators of the coset model must be invariant
under the BRST charge \eqref{brst}, they should take the
form
\begin{align}
 e^{i r H + u \chi} \Phi_{j,m}^s e^{  2 i b(m+r - \frac{u}{2}) X} ,
 \label{brstvo}
\end{align}
whose conformal weight is
\begin{align}
\Delta = - 2b^2j(j+\tfrac12) + \frac{s(s-1)}{2} + \frac{r^2}{2}
 - \frac{u^2}{2} + 2 b^2 \left(m+r-\frac{u}{2}\right)^2 .
\end{align}
Here we have used $\Phi_{j,m}^s$ as the vertex operator of
OSP(1$|$2) WZNW model as defined in \eqref{freev}.

Physical operators of the topological model can be constructed from
chiral primaries of the ${\cal N}=2$ coset model.
Here we review how to perform the topological twist to the
chiral primaries by following \cite{MV,Takayanagi}.
First we find chiral primary states of the coset in NS-sector,%
\footnote{Notice
that there are three different spin structures that appear in this
paper. One is for the OSP(1$|$2) current algebra, which is
defined such that an
an integer $s$ in \eqref{freev} means the NS-sector,
while a half integer $s$ implies R-sector.
The second spin structure is the
ordinary one for the $N=2$ superconformal field theory. The third
one is for the $\hat{c}\le 1$ superstring.
In this section the notion of NS,R is with respect to the second
spin structure.}
which satisfy
\begin{align}
 G^+_{r - 1/2} | { \rm NS } \rangle =  G^-_{r + 1/2} | { \rm NS } \rangle = 0
\end{align}
for $r = 0,1,\cdots$.
Among the vertex operators of the form \eqref{brstvo},
there are chiral primary operators ${\cal O}^{NS,s}_j$
corresponding to the above chiral primary states.
These chiral primaries can be mapped to
R-ground states by spectral flow operation.
Redefining the U(1)$_R$ current \eqref{supercon} as
\begin{align}
 J_R
 =  - 2 b^2 \hat J^3  - \psi^+ \psi^-
 -  \varphi^+ \varphi^- = - i b \partial X_R,
\end{align}
with $X_R(z) X_R(0) \sim \ln z$, the R-ground states are obtained by
${\cal O}^{R,s}_j =  e^{\frac{i}{2} b X_R} {\cal O}^{NS,s}_j$.
Finally, the elements of cohomology for the topological theory are
obtained by the topological twist as ${\cal
O}^{s=1}_{j} = e^{- \frac{i}{2} \sqrt{k} b X'_R}   {\cal
O}^{R,s}_{j}$.
Here we define $X'_R$ as
\begin{align}
 J_{R}' = - 2 J^3 - 3 \psi^+ \psi^- - 2 \varphi^+ \varphi^-
  + 2 ib (k-2) \partial X =: - i \sqrt{k} b \partial X_R '
\end{align}
from the expression of R-current \eqref{rtcurrent}.

Among the chiral primaries of the ${\cal N}=2$ coset model, we focus
on the following two types of operators in NS-sector;
\begin{align}
 {\cal O}^{NS,\frac{1}{2}}_j =
 e^{- \frac{1}{2} \chi} \Phi^{\frac{1}{2}}_{j,j-\frac{1}{4}} e^{2ibjX} , \qquad
 {\cal O}^{NS,1}_j = \Phi^{1  }_{j,j} e^{2ibjX} ,
 \label{cp}
\end{align}
which satisfy  $\Delta = q_R/2 = - j b^2 -1/4$
and $\Delta = q_R/2 = - j b^2$, respectively.
The R-sector ground states are constructed as
\begin{align}
 {\cal O}^{R,\frac{1}{2}}_j =  e^{\frac{i}{2} H}
 \Phi^{\frac{1}{2}}_{j,j-\frac{1}{4}} e^{2ib(j+\frac{1}{4})X} , \qquad
 {\cal O}^{R,1}_j = e^{\frac{i}{2} H + \frac{1}{2} \chi}
  \Phi^{1}_{j,j} e^{2ib(j+\frac{1}{4})X} .
 \label{rg}
\end{align}
After the topological twist we finally obtain
\begin{align}
 {\cal O}^{\frac{1}{2}}_j =  e^{ - i  H - \chi}
 \Phi^{\frac{1}{2}, w = 1}_{j,j-\frac{1}{4}} e^{2ib(j+\frac{1}{4 b^2})X} , \qquad
 {\cal O}^{1}_j = e^{ - i  H - \frac{1}{2} \chi}
  \Phi^{1,w=1}_{j,j} e^{2ib(j+\frac{1}{4 b^2})X} .
  \label{cpt}
\end{align}
Notice that vertex operators are spectrally
flowed in the sense of OSP(1$|$2) WZNW model as in \eqref{spectralflow}
during the procedure of topological twist.
Under the spectral flow action we may identify
$
\Phi_{j, j + \frac{s -1}{2} }^{s,w =  1}
 = \Phi_{-j-\frac{k}{2}+\frac{1}{4} ,
 j + \frac{s-1}{2} + \frac{k}{2}}^{s - \frac{1}{2}}
$.
Combining with the free field representation of vertex operators
\eqref{freev}, we find
\begin{align}
{\cal O}^{\frac{1}{2}}_j
  \sim c e^{ - \chi}
 e^{ 2ib(j+\frac{1}{4 b^2})X + 2 b (j + \frac{k}{2} - \frac14) \phi} ,
 \qquad
  {\cal O}^{1}_j = c e^{ - \frac{1}{2} \chi}
   e^{ \frac{i}{2} Y + 2ib(j+\frac{1}{4 b^2})X
   + 2 b ( j + \frac{k}{2} - \frac14) \phi} .  \label{phyo}
\end{align}
In the above, we renamed $c=\exp(-iH)$ as suggested by the previous
discussion. Moreover the $\beta$-ghost in the superstring
should be written as $\beta' = \partial \xi \exp(-\chi)$.
Therefore, we can say that these operators have
one $c$-ghost and picture $-1$.%
\footnote{The operators corresponding to those in the
other picture may be obtained by the action of operator similar
to the picture changing operator.}

Notice that the above two operators (\ref{phyo}) indeed coincide with the
tachyon and RR field vertex operators in the two dimensional type 0
superstring, respectively (see subsection \ref{secc}).
Actually they complete the list of physical operators since
there are no massive stringy modes in two dimensional superstring.
This fact may be seen by taking the light-cone gauge.
In this way, we have learned that the
physical states (chiral primary states) in the topological string on
OSP(1$|$2)/U(1) are mapped into the physical states in the two
dimensional type 0 string. We will study the relation between these
two theories in more detail below.

\section{OSP(1$|$2)/U(1) Coset from ${\cal N}=1$ Super Liouville}
\label{ribault}

In references \cite{RT,HS3} it was shown that arbitrary correlation
functions of primary fields in SL(2) WZNW model can be written in
terms of correlation functions of Liouville field theory.
This property may be useful to show the
equivalence between the scattering amplitudes in $c\leq 1$ bosonic
string and the topological string on $SL(2)/U(1)$.
The agreement for three-point functions between them has been shown in
\cite{Takayanagi}, and this is generalized by \cite{NN}
to arbitrary tree level amplitudes by utilizing the generalized
relation of \cite{Ribault}.
Recently it was shown in \cite{HS2} that correlation
functions of OSP(1$|$2) WZNW model can be written in terms of those
of ${\cal N}=1$ super Liouville field theory.
Later we would like to show the equivalence between
${\cal N} = 2$ coset model
of OSP(1$|$2)/U(1) and the $\hat c \leq 1$ superstring in the
level of amplitudes.  For the purpose we
generalize the relation such as to include
RR-sectors of fermions and spectrally flowed sectors of OSP(1$|$2)
model. In this section we derive the generalized relation in
the path integral formulation following \cite{HS3,HS2}.

\subsection{OSP(1$|$2) WZNW Model}

Let us start from the action of OSP(1$|$2) WZNW model.
In terms of free fields the action may be written as\footnote{
A derivation of this action can be found in \cite{HS2}.
This action hare is a bit different from the one in \cite{HS2},
but it is easy to see the equivalence between the two expressions.}
\begin{align}   \label{OSPaction}
 S^\text{WZNW} (g) = \frac{1}{2\pi} \int d^2 z
  \Bigl[ \frac12 \partial \phi  \bar \partial \phi
  + \frac{b}{8} \sqrt{g} {\cal R} \phi
  + \beta \bar \partial \gamma
  + \bar \beta \partial \bar \gamma
  + p \bar \partial \theta + \bar p \partial \bar \theta
  \\ \nonumber
  - \frac{1}{k}\beta \bar \beta e^{2b\phi }
   - \frac{1}{2k} (p + \beta \theta )
   ( \bar p + \bar \beta \bar \theta ) e^{b \phi } \Bigr] ,
\end{align}
where $\phi,\gamma,\bar \gamma , \theta , \bar \theta$
are related to the parameters of elements
$g \in \,$ OSP(1$|$2) and $\beta , \bar \beta , \theta
, \bar \theta$ are conjugate variables.
The generators of current algebra symmetry
are written as in \eqref{freefield}
in these variables.
Here we use the form of vertex operator as
\begin{align}
  V_j^{s,\bar s} (\mu | z ) =
  \mu^{j + \frac12 + \frac{s}{2}} \bar \mu^{j + \frac12 + \frac{\bar s}{2}}
 e^{i s Y  + i \bar s \bar Y }
 e^{\mu \gamma - \bar \mu \bar \gamma} e^{2 b (j + \frac{1}{2} )\phi } ~. \label{vmu}
\end{align}
For the NSNS-sector with $s , \bar s = 0 ,1$ these vertex operators
are the same as in \cite{HS2}.
The conformal weights are given as $\Delta =  - 2 b^2 j (j + 1/2)$.
The vertex operators in the RR-sector are given by spin fields
with $s = \bar s = 1/2$, and
the conformal weights are $\Delta =  - 2 b^2 j (j + 1/2) + 1/8$.
The above expression in so-called $\mu$-basis is useful for our purpose,
and it can be mapped to the $m$-basis expression given in \eqref{freev}
by%
\footnote{In order to compare with the previous notation, we may need
to perform a flip $j  \to -j - 1/2$. Moreover, it might be natural to
multiply the factor $
 N^{s.\bar s}_{j,m,\bar m} =
 \frac{\Gamma(-j + 1/2 -s/2 + m)}{\Gamma (j+ 1/2 + \bar s/2 - \bar m)}$
 as, e.g., in \cite{RT}; see also \cite{Giribet2006}. Here we remove it since it may diverge
 in our case.
}
\begin{align}
 \Phi^{s,\bar s}_{j,m,\bar m} =
  \int \frac{d^2 \mu}{|\mu|^2} \mu ^{ - m} \bar \mu^{ - \bar m}
V^{s,\bar s}_j (\mu | z) .
\end{align}
In some sense, the $\mu$-basis expression can be thought of
generating function of the $m$-basis expression.

Since the operators of topological model in \eqref{cpt} are
written in terms of OSP(1$|$2) vertex operators with spectral flow
index $w=1$, it is important to understand
the symmetry under the spectral flow.
The spectral flow action $\rho^w$ can be defined as
\begin{align}
 \rho^w ( J_n^3) = J^3_n - \frac{k}{2} w \delta_{n,0} , \qquad
 \rho^w (J_n^{\pm}) = J^{\pm}_{n \pm w}  , \qquad
 \rho^w (j_r^{\pm} ) = j^{\pm}_{r \pm \frac{w}{2}} ,
\end{align}
where the mode expansions are
$J^A (z) = \sum_n J^A_n z^{-n-1}$ with $A=\pm,3$ and
$j^\pm (z) = \sum_r j^\pm_r z^{-r-1}$.
We can easily see that the new currents satisfy the
same (anti-)commutation relations as before,
which implies that the spectral flow is the symmetry of
the current algebra OSP(1$|$2).
The vacuum state is defined such as
\begin{align}
\rho^w (J^A_n) | w \rangle = 0 , \qquad
\rho^w (j^{\pm}_r ) | w \rangle = 0
\end{align}
for $n,r \geq 0$. In terms of free fields, the vacuum state
  $|w \rangle = |w \rangle_{(\beta,\gamma)}
  \otimes  | w \rangle_{\phi  }
  \otimes  | w \rangle_{Y  } $ is characterized  as
\begin{align}
 \beta_{n - w} |w \rangle_{(\beta,\gamma)} = 0 , \qquad
 \gamma_{n+w} |w \rangle_{(\beta,\gamma)} = 0
 \label{cond1}
\end{align}
for $n \geq 0$, and moreover
\begin{align}
 | w \rangle_{\phi  } =
 e^{\frac{w}{2b} \phi }  | 0 \rangle_{\phi  } , \qquad
 | w \rangle_{Y  } = e^{ - \frac{i w}{2} Y }  | 0 \rangle_{Y  } .
 \label{cond2}
\end{align}
In the following we assume $w \geq 0$ and
denote $v^w (0)$ as the operator
corresponding to the state $|w \rangle$.

As discussed in \cite{Ribault,HS}, generic $N$-point functions
of operators with spectral flow can be reduced to
$N$-point functions
of \eqref{vmu} with the inversion of $v^w(\xi)$.
We choose the position of insertion $v^w (\xi)$ as $\xi = 0$
since it does not affect the following discussion.
In the path integral formulation they are given as
\begin{align}
 \left \langle \prod_{\nu = 1}^N
 V^{s_\nu , \bar s_\nu}_{j_\nu} (\mu_\nu | z_\nu ) v^w (0)
 \right \rangle
 = \int_{(w)} {\cal D} \phi {\cal D} ^2 \beta
{\cal D} ^2 \gamma {\cal D} ^2 \theta {\cal D} ^2 p
 e^{- S^\text{WZNW} (g) }
 \times \\ \times
  \prod_{\nu = 1  }^N
V^{s_\nu , \bar s_\nu } (\mu_\nu | z_\nu )
e^{w ( \phi (0)/2b - i Y (0) /2 ) } .
 \nonumber
\end{align}
The effects of insertion $v^w (0)$ appears in the right hand
side in two ways. One is the extra insertion of vertex operator
$e^{w ( \phi (0)/2b - i Y (0) /2 ) }$, and the other is the
restriction to the integration domain of $\beta , \bar \beta$
such that $\beta , \bar \beta$ have a zero of order $w$ at
$\xi = 0$. For more detail see \cite{HS}.

\subsection{OSP(1$|$2)--Super Liouville Correspondence}

Now that we have the OSP(1$|$2) WZNW model, we can derive
the relation between the correlation functions of OSP(1$|$2) WZNW
model and those of ${\cal N}=1$ super Liouville field theory
by following the analysis of \cite{HS2}.
For this purpose we first integrate $\beta,\gamma$ as in \cite{HS2}.
Integrations over $\gamma$ and $\bar \gamma$ lead to delta
functionals for $\beta$ and $\bar \beta$, which replace the fields
$\beta, \bar \beta$ by
\begin{align}
 \beta (x) = \sum_{\nu = 1}^N \frac{\mu_\nu}{x - z_\nu}
   = u \frac{x^w \prod_{i=1}^{N-2-w}(x-y_i)}
            {\prod_{\nu = 1}^N (x-z_\nu)}
   = : u {\cal B} (x) .
    \label{separation}
\end{align}
The insertion of $v^w(0)$ forces $\beta(x)$ to have a zero of order
$w$ at $x=0$ and this requirement gives constraints
\begin{align}
 \sum_{\nu = 1}^N \mu_\nu z_\nu^{- n} = 0
\end{align}
for $n = 0, 1, \cdots , w$. Since a 1-form with $N$ poles
on a sphere has $N-2$ zero's, $\beta$ can be represented as in the right hand
side by the positions of $N-2-w$ more zero's $y_i$.
In other words, the parameters $y_i$ are defined by the
equation \eqref{separation}, and the new parameters are
essential to relate the model to super Liouville theory
as seen below.
Moreover, we can see that the number of spectral flow $w$
is restricted as $ w \leq N - 2$.

After the integration over $\beta,\gamma$,
the action becomes something similar to super Liouville theory,
but the coefficients  include functions ${\cal B} (z), \bar {\cal B}(\bar z)$.
Following \cite{HS2,HS} these can be removed by the
redefinition of fields as
\begin{align}
 \phi ' : = \phi  + \frac{1}{2 b} \ln | u {\cal B } | ^2 ,
 \qquad Y ' : =  Y  - \frac{i}{2} \ln | u {\cal B } | ^2 .
\end{align}
Moreover, after some manipulations we find the relation
\begin{align}
 & \left \langle \prod_{\nu = 1}^N
 V^{s_\nu,\bar s_\nu}_{j_\nu} (\mu_\nu | z_\nu)
  v^{w}(0)\right\rangle  \nonumber \\
  & \qquad = \prod_{n=0}^w \delta ^2 (\sum_\nu \mu_\nu z^{-n} )
  |u|^{2-\frac{w}{2b^2}+\frac{w}{2}} | \Theta^w_N |^2
 \left\langle \prod_{\nu = 1}^N
  V^{s_\nu-\frac{1}{2},\bar s_\nu - \frac{1}{2}}_{\alpha_\nu} ( z_\nu)
   \prod_{j=1}^{N-2-w} V^{\frac{1}{2},\frac{1}{2}}_{-\frac{1}{2b}} (y_j) \right\rangle
   \label{Ribault}
\end{align}
with $\alpha_\nu = 2b (j_\nu + 1/2 ) +1/2b $.
The right hand side is computed by the sum of ${\cal N}=1$
super Liouville theory $(\phi ',\psi,\bar \psi)$  and massless fermions $(\psi_X,\bar \psi_X)$
\begin{align}
   S [\phi ',\psi,\psi_X]
  & = \  \frac{1}{4\pi} \int d^2 z \, \Bigl[\,  \partial \phi ' \bar \partial \phi '
   + \frac{Q_{\phi ' }}{4} \sqrt g R \phi ' + \frac{2}{k} e^{ 2 b \phi '}\, +
   \nonumber \\[2mm]
 & \hspace*{2cm} +
   \psi \bar \partial \psi + \bar \psi \partial \bar \psi
 + \psi_X \bar \partial \psi_X + \bar \psi_X \partial \bar \psi_X
 - \frac{2}{k} \psi \bar \psi e^{ b \phi '}\,  \Bigr]
 \label{sLaction}
\end{align}
with $Q_{\phi ' } = b + 1/b$.
The fermions are defined by
\begin{align}
 \psi \pm i \psi_X = \sqrt{2} e^{ \pm i Y '} , \qquad
 \bar  \psi \pm i \bar \psi_X = \sqrt{2} e^{ \pm i \bar Y '} ,
\end{align}
and the vertex operators are
\begin{align}
V^{s,\bar s}_{\alpha} ( z )
 = e^{i s Y + i s \bar Y}
   e^{\alpha \phi '}
\end{align}
with conformal weights $\Delta = \alpha ( Q_{\phi ' } - \alpha )/2
 + s^2 /2 $.
The twist factor is
\begin{align}
 \Theta^{w}_N = \prod_{\mu < \nu}^N (z_{\mu} - z_{\nu})
  ^{\frac{1}{4b^2} -\frac{1}{4}}
   \prod_{i<j}^{N-2-w} (y_{i} - y_{j})^{\frac{1}{4b^2} - \frac{1}{4}}
    \prod_{\nu=1}^N \prod_{i=1}^{N-2-w} (z_\nu - y_j)^{-\frac{1}{4b^2} + \frac{1}{4}} .
\end{align}
Here we should notice that the operators in the NSNS(RR)-sector are mapped
to those in the RR(NSNS)-sector. Moreover, if the winding number
is violated maximally as $w = N -2$, then there is no extra insertion of
operator at $z = y_i$.

\subsection{Amplitudes of OSP(1$|$2)/U(1) Coset Model}

Utilizing the fundamental relation \eqref{Ribault}, we can
rewrite correlation functions of OSP(1$|$2)/U(1)
coset in terms of ${\cal N}-1$ super Liouville theory with a supersymmetric pair of free boson and fermion
in a manner similar to the bosonic case \cite{HS}.
Here the vertex operators of coset model are defined as
in \eqref{parafermion}%
\footnote{Only in this subsection we construct the coset model by gauging
the axial U(1) symmetry in order to use the trick of \cite{HS}.
}
\begin{align}
 \Psi^{s,\bar s}_{j,m,\bar m} (z,\bar z)  =
 V^{X^3}_{m , \bar m} (z , \bar z)
 \Phi^{s,\bar s}_{j,m,\bar m}  (z , \bar z)
\end{align}
with
\begin{align}
 V^{X^3}_{m,\bar m} (z,\bar z) =
 e^{i \sqrt{\frac{2}{k}}( - m X^3 + \bar m \bar X^3)} .
\end{align}
Moreover, the vertex operators of OSP(1$|$2) model
with spectral flow index $w$
are related to vertex operators of OSP(1$|$2)/U(1) model as
(see \eqref{spectralflow})
\begin{align}
 \Psi^{s,\bar s}_{j,m,\bar m} (z,\bar z)  =
 V^{X^3}_{m + \frac{kw}{2}, \bar m + \frac{kw}{2}} (z , \bar z)
 \Phi^{s,\bar s , w}_{j,m,\bar m}  (z , \bar z) .
 \label{sf2}
\end{align}
Therefore we can also obtain a formula for correlation functions
of OSP(1$|$2) model with non-trivial spectral flow actions.
In particular, we will
be interested in a specific $N$-point function in OSP(1$|$2) WZNW model
as
\begin{align}
 {\cal M} = \left \langle \Phi^{s_1,\bar s_1}_{j_1,m_1,\bar m_1}  (z_1)
 \Phi^{s_2,\bar s_2}_{j_2,m_2,\bar m_2}  (z_2)
\prod_{\nu = 3}^N
 \Phi^{s_\nu ,\bar s_\nu , w_\nu = 1}_{j_\nu,m_\nu,\bar m_\nu}  (z_\nu )
 \right  \rangle .
 \label{MWV0}
\end{align}
Since the amplitude has $N-2$ number of winding violation, it should be
written in terms of $N$-point function of ${\cal N}=1$ super Liouville theory.

Let us first study $N$-point function of OSP(1$|$2)/U(1) coset model.
As before we introduce a new field $\hat X^3$ by
\begin{align}
 \hat X^3_L : = X^3_L - i \sqrt{\frac{k}{2}}
\ln  ( u {\cal B} ) ,
 \end{align}
and the right mover defined by its complex conjugate.
By closely following \cite{HS3} and utilizing the
formula \eqref{Ribault}, we finally obtain
\begin{align} \label{cigar}
&  \left \langle \prod_{\nu = 1}^N
 \Psi^{s_\nu , \bar s_\nu}_{j_\nu, m_\nu , \bar m_\nu}
 \right  \rangle =
  \int \frac{\prod_{i=1}^{N - 2 - w} d ^2 y_i}{(N-2-w)!}
 \times \\
& \qquad \qquad \times
 \left  \langle \prod_{\nu=1}^N
 V^{\hat X^3}_{m_\nu + \frac{k}{2}, \bar m_\nu + \frac{k}{2}} (z_\nu)
  V^{s_\nu - \frac{1}{2}, \bar s_\nu- \frac{1}{2}}_{\alpha_\nu} (z_\nu  )
  \prod_{i =1}^{N-2 - w} V^{\hat X^3}_{- \frac{k}{2},- \frac{k}{2}}(y_i)
 V^{\frac{1}{2},\frac{1}{2}}_{- \frac{1}{2b}} (y_i  )
 \right  \rangle .
  \nonumber
\end{align}
The label $w$ is related to the winding number violation as
$\sum_\nu m_\nu = \sum_\nu \bar m_\nu = - \frac{kw}{2}$.
The right hand side should be computed by the theory
with the action $S[\phi ',\psi,\psi_X]$ for ${\cal N} = 1$ super
Liouville theory and a free fermion $(\psi_X, \bar \psi_X)$
and a free boson $\hat X^3$ with background charge
$Q = - i \sqrt{k}$ for its dual field.

With the formula \eqref{cigar} and \eqref{sf2}
we can write down
generic correlation functions in the OSP(1$|$2) model
with spectral flow action considered
in terms of super Liouville theory.
Here we only compute the amplitude \eqref{MWV0} since it
is the case  used in the later analysis.
With the formula \eqref{sf2} we can relate the amplitude
\eqref{MWV0} to a $N$-point function of the coset as
\begin{align}
 \left \langle \prod_{\nu = 1}^N
 \Psi^{s_\nu , \bar s_\nu}_{j_\nu, m_\nu , \bar m_\nu}
 \right  \rangle & = {\cal M} \times \left \langle
 V^{X^3}_{m_1 , \bar m_1 } (z_1 )
 V^{X^3}_{m_2 , \bar m_2 } (z_2 )
\prod_{\nu = 3}^N
 V^{X^3}_{m_\nu + \frac{k}{2}, \bar m _\nu + \frac{k}{2}} (z_\nu )
\right  \rangle .
\end{align}
Then, by combining with the formula \eqref{cigar}, we find
\begin{align}
{\cal M} = |\Theta_s (z_\nu )|^2
 \left  \langle \prod_{\nu=1}^N
  V^{s_\nu - \frac{1}{2}, \bar s_\nu- \frac{1}{2}}
  _{\alpha_\nu} (z_\nu  )
 \right  \rangle ,
 \label{MWV}
\end{align}
where the right hand side is computed by the ${\cal N}=1$
super Liouville theory and a free fermion with the action
\eqref{sLaction}. The coefficient is given by
\begin{align}
\Theta_s (z_\nu ) = (z_1 - z_2)^{\frac{k}{2} + m_1 + m_2 }
 \prod_{\nu = 3}^N [ (z_1 - z_\nu ) (z_2 - z_\nu ) ]
 ^{\frac{k}{2} + m_\nu} ~,
\end{align}
and bared expression for $\bar \Theta_s (\bar z_\nu )$.
This formula will be important to relate amplitudes of
the topological model and the $\hat c \leq 1$ superstring theory.

\section{Correspondence to $\hat c \leq 1$ Superstring Theory}

In section \ref{coset}, we have studied the topological model
based on the ${\cal N}=2$ supersymmetric coset OSP(1$|$2)/U(1).
In particular, we have observed that free fields and chiral primaries
of the coset model are mapped to matter contents and physical
operators in the $\hat c \leq 1$ superstring theory.
In this section, we establish the relation in more detail.
After briefly reviewing the $\hat c \leq 1$
superstring theory and the method to compute
amplitudes in topological models, we compare
the amplitudes of both theories.

\subsection{$\hat c \leq 1$ Superstring Theory}
\label{secc}

In section \ref{coset} we have already encountered the
$\hat c \leq 1$ superstring theory during constructing the
topological model of the coset OSP(1$|$2)/U(1).
In this subsection we define the $\hat c \leq 1$ superstring
theory in a more precise way.%
\footnote{Notice that we are setting $\alpha'=2$ in this
paper.} The matter part consists of a linear dilaton $X$ with
background charge $Q_X = i(1/b - b)$ and a free fermion $\psi_X$.
The action of these fields is given by
\begin{align}
 S_X = \frac{1}{4 \pi} \int d^2 z \left[ \partial X \bar \partial X
  + \frac{Q_X}{4}\sqrt{g} {\cal R} X + \psi_X \bar \partial \psi_X
  + \bar \psi_X \partial \bar \psi_X \right] .
\end{align}
The theory also includes the ${\cal N } = 1$ super Liouville theory,
whose action is
\begin{align}
 S = \frac{1}{4 \pi} \int d^2 z \left[ \partial \phi  \bar \partial \phi
  + \frac{Q_{\phi } }{4}\sqrt{g} {\cal R} \phi + \psi \bar \partial \psi
  + \bar \psi \partial \bar \psi + \mu_L \psi \bar \psi e^{b \phi}
 \right]
\end{align}
with $Q_\phi = b + 1/b$.
The total central charge is now
$c = 3/2 + 3 Q_X^2 + 3/2 + 3 Q_\phi^2 = 15$,
and hence we can construct a critical superstring theory by
coupling the world-sheet superghosts
$(b,c)$ and $(\beta ',\gamma ')$.

Primary operators of this theory may take the form
$\exp ( \alpha X + \beta \phi)$, which
has the conformal weight $\Delta = \alpha (
Q_X - \alpha ) /2 +  \beta ( Q_\phi - \beta ) /2 $.
Following the standard method to construct BRST invariant
operators, we can find out physical operators in
the $\hat c \leq 1$ superstring theory.
The tachyon vertex operator is given as
\begin{align}
c \bar c {\cal T}^{(-1)}_p = c \bar c e^{- ( \chi + \bar \chi ) }
e^{i k_X (X + \bar X) + k_\phi^\pm  \phi  } ,
\label{m1picture}
\end{align}
where the momenta run over
$ i k_X = Q_X /2 + i p$ and $k_\phi^\pm = Q_\phi / 2 \pm p$
with $p \in \mathbb{R}$.
We bosonize the superghost $\beta ' , \gamma '$ like in
\eqref{bosonized}, which yields the new fields $\chi$.
In other words, the above expression is in
$(-1,-1)$ picture; and
in $(0,0)$ picture it is written as
\begin{align}
c \bar c {\cal T}^{(0)}_p = c \bar c ( i k_X \psi_X  + k_\phi^\pm \psi  )
  ( i k_X \bar \psi_X  + k_\phi^\pm \bar \psi  )
 e^{ i k_X (X + \bar X ) + k_\phi^\pm \phi } .
 \label{0picture}
\end{align}
There are other physical operators in the RR-sector.
The Ramond vertex operator in $(-1/2 , -1/2)$ picture
is written as
\begin{align}
 c \bar c {\cal R}^{(-1/2)}_p = c \bar c
 e^{- \frac12 ( \chi + \bar \chi ) }
 e^{  \pm \frac{i}{2} (Y + \bar Y ) + i k_X (X_L + X_R ) + k^\pm_\phi \phi } .
 \label{rrv}
\end{align}
Indeed, these vertex operators (\ref{m1picture}) and (\ref{rrv}) are
the same as those obtained from the topological string on
OSP(1$|$2)/U(1) as observed in (\ref{phyo}).

If $X$ direction is compactified with radius $R$, then the momentum
takes discrete values $p = n /R$ with $n \in {\mathbb Z}$.
For winding modes we should replace
$X_R \to - X_R$, $\bar H \to - \bar H$ and $p = w R/2$ with $w \in {\mathbb Z}$.
After the proper GSO projection, type 0B theory has the tachyon modes and the
momentum modes in the RR-sector, On the other hand, type 0A theory has the tachyon
modes and the winding modes in the RR-sector.
See, e.g. \cite{TT,DKKMMS} for more detail.

\subsection{Amplitudes of Topological Model}
\label{atm}

Before dealing with the specific case of OSP(1$|$2)/U(1),
we give generic arguments on
amplitudes in topological models.
Let us consider a topological field theory obtained by the
topological twist $T^{top} (z) = T (z) + 1/2 \partial J_R (z) $ of a
${\cal N} = 2$ super conformal field theory \cite{EY,WTST}. We
consider the B model, namely twist the same way for the
anti-holomorphic part as $\bar T^{top} (\bar z) = \bar T (\bar z) +
1/2 \bar \partial \bar J_R (\bar z) $. Then the physical spectrum
can be computed from the cohomology of BRST operator $Q = \oint G^+
(z) dz$. Let us write a basis of physical operators (in NS sector)
as $\phi_i$, then we can obtain other types of physical operators as
\begin{align}
 \oint dz G^-_{- \frac12} \cdot \phi_i , \qquad
 \oint  d \bar z \bar G^-_{- \frac12} \cdot \phi_i , \qquad
 \int d^2 z G^-_{- \frac12} \bar  G^-_{- \frac12} \cdot \phi_i .
\end{align}
Following the arguments on \cite{DVV2,WTST,BCOV},
we compute amplitudes of the form
\begin{align}
 {\cal F}
 =  \left \langle \phi_{i_1} (z_1 )
 \phi_{i_2} (z_2 ) \phi_{i_3} (z_3 )
 \prod_{\nu =4}^N \left [ \int d ^2 z_\nu \tilde \phi_{i_\nu}(z_\nu) \right ]   \right \rangle ,
 \label{topamp}
\end{align}
which would give us
non-trivial information of the topological model.
Here we have defined
$\tilde \phi_i = G^-_{-\frac{1}{2}} \bar G^-_{-\frac{1}{2}} \phi_i$.

An important fact is that
the above amplitudes of topological model can be computed in
the original untwisted model \cite{WTST}. After the topological
twist the $U(1)_R$ current becomes anomalous, and hence we should
insert $U(1)_R$ fields into correlators of original model to reproduce
the topological amplitudes. Here we insert the operator
$\mu (z ,
\bar z) = e^{ \frac{i}{2} \sqrt{\frac{c}{3}}( X _R (z)) + \bar X _R
( \bar z))}$ at two points $z = z_1,z_2$, where
the free boson $X_R$ is related to the R-current as
$J_R (z) = - i
\sqrt{\frac{c}{3}} \partial X _R (z)$. This operator maps the
physical operators $\phi_i$ of the topological model to operators
$\phi_i^R$ in the R-ground states of the original model.
In this way, we can write the topological amplitude \eqref{topamp} as
\begin{align} \label{orgamp}
 {\cal F}
 &= |z_1 - z_2 |^{q_1 + q_2}
 ( |z_1 - z_3 | |z_2 - z_3 | )^{q_3}
  \times \\ &\times \nonumber
   \left \langle \phi^R_{i_1} (z_1 )
 \phi^R_{i_2} (z_2 ) \phi_{i_3} (z_3 )
 \prod_{\nu =4}^N \left [ \int d ^2 z_\nu
 ( | z_1 - z_\nu | | z_2 - z_\nu | )^{q_\nu - 1}
 \tilde \phi_{i_\nu}(z_\nu) \right ]   \right \rangle  ,
\end{align}
where the right hand side is computed in the original model
before the topological twisting.
Here we have denoted
$q_\nu$ as the $U(1)_R$ charge of $\phi_{i_\nu}$.

\subsection{Comparison of Correlation Functions}
\label{correlation}

After the preparation we can now compare correlation functions
of topological model on OSP(1$|$2)/U(1) and of the $\hat c \leq 1$
superstring. We start from the amplitude of the topological
model, and then show the equivalence by using the formula
\eqref{MWV} obtained above.
 Here we
only consider the amplitudes of operator of the first type in
\eqref{cpt}, which corresponds to the tachyon operator
in the $\hat{c} \leq 1$ superstring. The case with the second type
in \eqref{cpt} can be analyzed in a similar way.

Since the conservation of U(1) current $J
=\partial \chi$ is violated by the amount of $-2$, non vanishing
amplitudes may be given as
\begin{align}
 {\cal F}
 =  \left \langle {\cal O}^{\frac12}_{j_1} (z_1 )
 {\cal O}^{\frac12}_{j_2} (z_2 ) {\cal O}^{-\frac12}_{j_3}(z_3 )
 \prod_{\nu =4}^N \left [ \int d ^2 z_\nu \tilde
  {\cal O}^{-\frac12}_{j_\nu} (z_\nu) \right ]   \right \rangle .
  \label{topamp2}
\end{align}
This violation corresponds to the fact that the sum of picture
must be $-2$ in superstring theory.
The operator ${\cal O}^{\frac{1}{2}}_j$ is defined in
\eqref{cpt} as
\begin{align}
 {\cal O}^{\frac{1}{2}}_j  =  e^{ - i  (H + \bar H ) - ( \chi + \bar \chi ) }
 \Phi^{\frac{1}{2},\frac{1}{2}, w = 1}
    _{j,j-\frac{1}{4},j-\frac{1}{4}}
 e^{2ib(j+\frac{1}{4 b^2}) (X + \bar X )} .
\end{align}
Following the argument in the previous subsection, this
operator would be mapped to the R-ground state operator of
the original model
\begin{align}
 {\cal O}^{R,\frac{1}{2}}_j =  e^{\frac{i}{2} (H + \bar H )}
 \Phi^{\frac{1}{2},\frac{1}{2}}
 _{j,j-\frac{1}{4},j-\frac{1}{4}} e^{2ib(j+\frac{1}{4})(X+ \bar X)} .
\end{align}
Another operator $ {\cal O}^{- \frac{1}{2}}_j$
is given by a linear combination of
\begin{align}
  e^{ - i  (H + \bar H ) }
 \Phi^{s, \bar s, w = 1}
   _{j,j + \frac{1}{4},j+\frac14} e^{2ib(j+\frac{1}{4 b^2})(X+ \bar X)}
\end{align}
with $s , \bar s = -1/2 ,3/2 $.
This operator should correspond to picture $(0,0)$ tachyon, and
can be constructed
by following the analysis in section \ref{coset}.
The other operator $ \tilde {\cal O}^{- \frac{1}{2}}_j$
is then generated by the action of $G^-_{-1/2} \bar G^{-}_{-1/2}$
as mentioned before
and written as a linear combination of
\begin{align}
 \Phi^{s, \bar s, w = 1}
   _{j,j - \frac{3}{4},j-\frac34} e^{2ib(j+\frac{1}{4 b^2})(X+ \bar X)}
\end{align}
with $s , \bar s = -1/2 ,3/2 $.

As argued in the previous subsection,
we first map the amplitude of topological model
\eqref{topamp2} to that of original model before twisting as in
\eqref{orgamp}.
Then we can use the formula \eqref{MWV} to relate it to the
amplitude of ${\cal N}=1$ super Liouville theory.
Following the analysis of \cite{NN,Takayanagi} we can then show that
\begin{align}
 {\cal F}
 =
 \left \langle c \bar c {\cal T}^{(-1)}_{j_1} (z_1 )
 c \bar c {\cal T}^{(-1)}_{j_2} (z_2 )
 c \bar c {\cal T}^{(0)}_{j_3}(z_3 )
 \prod_{\nu =4}^N \left [ \int d ^2 z_\nu
  {\cal T}^{(0)}_{j_\nu} (z_\nu) \right ]   \right \rangle
\end{align}
up to some coefficients.
Here, the operators ${\cal T}^{(p)}$ are tachyon operators
in the $p$-th picture and they are given by
\begin{align}
&{\cal T}^{(-1)}_p = e^{- ( \chi + \bar \chi ) }
e^{i k_X (X + \bar X) + k_\phi^+ \phi  } ~, \\
&{\cal T}^{(0)}_p =  ( i k_X \psi_X  + k_\phi^+ \psi  )
  ( i k_X \bar \psi_X  + k_\phi^+ \bar \psi  )
 e^{ i k_X (X + \bar X ) + k_\phi^+ \phi }
\end{align}
with $k_X = (1/b - b)/2 + 2b ( j + 1/4)$ and
$k_\phi^+ = (1/b + b )/2 + 2b ( j + 1/4 )$.
In this way we have shown that the amplitude of topological
string on OSP(1$|$2)/U(1) can be identified with that of
the $\hat c \leq 1$ superstring.

\section{Conclusion and Discussions}
\label{conclusion}

In this paper we have proposed an equivalence between the
topological string theory based on the coset OSP(1$|$2)/U(1)
and the $\widehat{c}\leq 1$
superstring theory. The latter is constructed by coupling a
$\hat{c}\leq 1$ matter to the $\mathcal{N}=1$
super Liouville theory. This can be regarded as a
supersymmetric version of the equivalence between the topological
string on $SL(2)/U(1)$ and the $c\leq 1$ bosonic string, which was
originally discovered by Mukhi and Vafa \cite{MV} for the case $c=1$ and was
later generalized to the $c<1$ case in \cite{Takayanagi}.
First we showed in the free field description that the field contents and the physical operators of the world-sheet theories of both string theories
match. Moreover, we
investigated the proposed equivalence at the level of scattering amplitudes
by applying the map \cite{HS2} between correlation functions in the
OSP(1$|$2) WZNW model and in super Liouville field theory.

This map
is a supersymmetric version of the one found by Ribault and Teschner
to relate correlation functions in the SL(2) WZNW model and those in
Liouville theory \cite{RT,HS3}. In the last years, the result has been
used with great success to investigate different dualities between
non-rational conformal field theories. In particular, it has led to the
proof of Fateev-Zamolodchikov-Zamolodchikov conjecture in \cite{HS}.
The duality between different non-rational two-dimensional conformal field
theories has a long story, and now a considerable list of examples is
available: quantum Hamiltonian reduction \cite{BO2},
Mukhi-Vafa duality \cite{MV,Takayanagi}%
, and Fateev-Zamolodchikov-Zamolodchikov duality
\cite{FZZunpublished} (and its supersymmetric version \cite{HoKa})
are probably the most renowned examples. These examples were, in
fact, very useful to study string theory. For instance, it was the
Fateev-Zamolodchikov-Zamolodchikov duality what really permitted to
construct a dual matrix model for strings in the the 2D black hole
background \cite{KKK}. It is our hope that the new equivalence
between conformal theories we studied in this paper will be relevant
to understand new aspects of superstring dualities as well.

There are a number of issues which should be understood in the future.
First we would like to understand better the spin structure and the
picture changing operation of the topological string theory on the
supercoset. It is also important to prove the complete equivalence
of physical states between these two theories. An exhaustive analysis of the cohomology of the theory is needed to this end. Finally, it would be
nice if we could understand a geometrical interpretation of the
supercoset OSP(1$|$2)/U(1) in terms of a certain
(maybe super) Calabi-Yau manifold, as SL(2)/U(1) coset
model is related to the conifold.

\subsection*{Acknowledgement}

We are very grateful to K. Hori for a useful discussion. GG thanks
the members of the IPMU for their hospitality during his stay.
YH would like to thank T.~Creutzig, H.~Irie and P.~B.~R{\o}nne for
useful discussions. The
work of GG has been partially supported by ANPCyT grant
PICT-2007-00849, by UBACyT grants X861 and X432, and by JSPS-CONICET
cooperation programme.
The work of YH is supported by JSPS Research Fellowship.
TT is supported by World Premier
International Research Center Initiative (WPI Initiative), MEXT,
Japan.
 The work of TT is supported in part by JSPS Grant-in-Aid for
Scientific Research No.20740132 and by JSPS Grant-in-Aid for
Creative Scientific Research No. 19GS0219.


\appendix


\section{Free Field Correlation Functions}

\label{ap:free}

Although the computation of $N$-point functions in the OSP(1$|$2) WZNW model
also involves fermion contributions and the insertion of picture changing
operators, the building blocks to construct such observables are the
correlation functions of vertices (\ref{freev}). Let us discuss these
correlation functions in the free field representation proposed here.
Consider the vertex operators
\begin{equation}
\Phi _{j,m}^{s}(z)=N_{j,m}^{s}\ \gamma _{(z)}^{-j-1/2+m-s/2}e^{2(j+1/2)b\phi
(z)}e^{isY(z)}\times h.c.  \label{Vertice}
\end{equation}%
where $h.c.$ stands for the anti-holomorphic portion of the operator,%
\footnote{More precisely, the $h.c.$ refers to the "bared contribution", and not
necessarily to the anti-holomorphic part. Actually, labels $s,m$ and $\bar{s}%
,\bar{m}$ are not necessarily related by complex conjugation.} and $%
N_{j,m}^{s}$ refers to the normalization.

Correlation functions are defined as follows;
\begin{equation*}
\left\langle \prod\nolimits_{\nu =1}^{N}\Phi _{j_{n},m_{\nu }}^{s_{\nu
}}(z_{\nu })\right\rangle =\int \mathcal{D}\phi \mathcal{D}^{2}\beta
\mathcal{D}^{2}\gamma \mathcal{D}^{2}\theta \mathcal{D}^{2}p\ e^{-S^{\text{%
WZNW}}(g)}\prod\nolimits_{\nu =1}^{N}\Phi _{j_{n},m_{\nu }}^{s_{\nu
}}(z_{\nu })
\end{equation*}%
where $S^{\text{WZNW}}(g)$ refers to the action of the WZNW model
(\ref{OSPaction}). It is convenient to consider again the
bosonization (\ref{Bosonization}); that is, defining\ $\theta =e^{iY}$, $%
p=e^{-iY}$. The existence of non-trivial background charges associated to
the fields $\phi $ and $Y$ requires special treatment of correlators. As
usual in the Coulomb gas representation, this charge compensation is
achieved by inserting additional operators that contribute to screen the
charges at infinity. Screening operators are exact marginal deformations of
the affine theory. In this theory four operators of this kind are available;
namely%
\begin{eqnarray}
\mathcal{S}_{++}(z,\bar{z}) &=&\frac{\lambda }{2k}S_{+}(z)\bar{S}_{+}(\bar{z}%
),\qquad \mathcal{S}_{+-}(z,\bar{z})=\frac{\lambda }{2k}S_{+}(z)\bar{S}_{-}(%
\bar{z}),  \label{CurlyS1} \\
\mathcal{S}_{-+}(z,\bar{z}) &=&\frac{\lambda }{2k}S_{-}(z)\bar{S}_{+}(\bar{z}%
),\qquad \mathcal{S}_{--}(z,\bar{z})=\frac{\lambda }{2k}S_{-}(z)\bar{S}_{-}(%
\bar{z}),  \label{CurlyS}
\end{eqnarray}%
with%
\begin{equation}
S_{+}(z)=e^{-iY(z)}e^{b\phi (z)},\qquad S_{-}(z)=\beta
_{(z)}e^{+iY(z)}e^{b\phi (z)},  \label{CurlySS}
\end{equation}%
and where $\lambda $ is a constant (see below).

\thinspace
The $N$-point correlation functions are thus defined by inserting
different amount of screening operators (\ref{CurlyS1})-(\ref{CurlyS}) in
the correlators, in addition to the $N$ vertex operators. Non-vanishing
correlation functions are given by
\begin{eqnarray}
n_{+}-n_{-} &=&\sum\nolimits_{\nu =1}^{N}s_{\nu }-1,\qquad \bar{n}_{+}-\bar{n%
}_{-}=\sum\nolimits_{\nu =1}^{N}\bar{s}_{\nu }-1  \label{Uno1} \\
n_{+}+n_{-} &=&\bar{n}_{+}+\bar{n}_{-}=-2\sum\nolimits_{\nu =1}^{N}j_{\nu
}+1-N  \label{Dos2}
\end{eqnarray}%
together with
\begin{equation}
\sum\nolimits_{\nu =1}^{N}m_{\nu }=\sum\nolimits_{\nu =1}^{N}\bar{m}_{\nu
}=0,  \label{Tres3}
\end{equation}%
where $n_{\pm }$ (and $\bar{n}_{\pm }$) are the amount of operators of the
type $S_{\pm }(z)$ (resp. $\bar{S}_{\pm }(\bar{z})$) in the correlators.
Equations (\ref{Uno1})-(\ref{Tres3}) determine the amount of screening
operators in terms of the quantum number of the vertices involved in the
correlators.

To illustrate the Coulomb gas prescription, let us consider the sector $%
s_{\nu }=\bar{s}_{\nu }$, which yields $n_{\pm }=\bar{n}_{\pm }$. In this
case, correlation functions are  given by contributions of the form
\begin{equation}
\frac{(\lambda /2k)^{n_{+}+n_{-}}}{n_{+}!n_{-}!}%
\int
\prod\nolimits_{r=1}^{n_{+}}d^{2}w_{r}\prod\nolimits_{l=1}^{n_{-}}d^{2}y_{l}%
\left\langle \prod\nolimits_{\nu =1}^{N}\Phi _{j_{n},m_{\nu
}}^{s_{\nu }}(z_{\nu })\prod\nolimits_{r=1}^{n_{+}}\mathcal{S}%
_{++}(w_{r})\prod\nolimits_{l=1}^{n_{-}}\mathcal{S}_{--}(y_{l})\right\rangle
_{\text{free}}  \label{B2}
\end{equation}%
where the subscript "$\mathrm{free"}$ refers to the fact that this correlator
is defined in the free field theory. The amount of screening insertions
$n_{\pm}$ in (\ref{B2}) is given by (\ref{Uno1}) and (\ref{Dos2}). Correlators similar to (\ref{B2})
but with a different amount of screening insertions $\prod\nolimits_{r=1}^{n_{+}-n} \mathcal{S}_{++}(w_{r})$
$\prod\nolimits_{l=1}^{n_{-}-n}\mathcal{S}_{--}(y_{l})$
$\prod\nolimits_{t=1}^{n} \mathcal{S}_{-+}(\hat{w}_{t})$
$\prod\nolimits_{s=1}^{n}\mathcal{S}_{+-}(\hat{y}_{s})$
also contribute. All the
contributions are gathered with an appropriate prescription to integrate the screening operators in the world-sheet.

The inclusion of screening operators (\ref{CurlyS1})-(\ref{CurlyS}) in (\ref%
{B2}) can be also thought of as coming from the interaction terms in the
action $S^{\text{WZNW}}(g)$. In this picture, conditions (\ref{Uno1})-(%
\ref{Tres3}) emerge from the integration over the zero-modes of the fields.
The scale $\lambda $ is easily introduced by shifting the zero mode of $\phi
$ as $\phi (z)\rightarrow \phi (z)+b^{-1}\log (\lambda )$. The parameter $%
\lambda $ allows to keep track of the KPZ scaling of correlation functions
\cite{KPZ}.

Correlation functions (\ref{B2}) can be computed by using free field
propagators (\ref{Propagators}),
\begin{eqnarray}
\left\langle \prod\nolimits_{\nu =1}^{N}\Phi _{j_{n},m_{\nu }}^{s_{\nu
}}(z_{\nu })\right\rangle  &=&\prod\nolimits_{\nu =1}^{N}N_{j_{\nu },m_{\nu
}}^{s_{\nu }}\frac{(\lambda /2k)^{n_{+}+n_{-}}}{n_{+}!n_{-}!}\int
\prod\nolimits_{r=1}^{n_{+}}d^{2}w_{r}\prod\nolimits_{l=1}^{n_{-}}d^{2}y_{l}
\notag \\
&&\times \left\langle \prod\nolimits_{\nu =1}^{N}e^{is_{\nu }Y(z_{\nu
})}\prod\nolimits_{r=1}^{n_{+}}e^{-iY(w_{r})}\prod%
\nolimits_{l=1}^{n_{-}}e^{iY(y_{l})}\right\rangle _{\text{free}}\times
\notag \\
&&\times \left\langle \prod\nolimits_{\nu =1}^{N}e^{b(2j_{\nu }+1)\phi
(z_{\nu })}\prod\nolimits_{r=1}^{n_{+}}e^{b\phi
(w_{r})}\prod\nolimits_{l=1}^{n_{-}}e^{b\phi (y_{l})}\right\rangle _{\text{%
free}}\times   \notag \\
&&\times \left\langle \prod\nolimits_{\nu =1}^{N}\gamma _{(z_{\nu
})}^{m_{\nu }-j_{\nu }-(s_{\nu }+1)/2}\prod\nolimits_{l=1}^{n_{-}}\beta
_{(y_{l})}\right\rangle _{\text{free}}\times h.c.
\end{eqnarray}%
where, again, $N_{j,m}^{s}$ is the normalization of the vertex. The standard
normalization $N_{j,m}^{s}=\frac{\Gamma (-j+1/2-s/2+m)}{\Gamma (j+1/2+\bar{s}%
/2+\bar{m})}$ yields
\begin{equation*}
\left\langle \Phi _{j,m}^{s}(z_{1})\Phi
_{-j-1/2,-m}^{1-s}(z_{2})\right\rangle = |z_{1}-z_{2}|^{-4\Delta }.
\end{equation*}
By expanding this expression, after Wick contracting all the contributions,
it takes the form%
\begin{equation*}
\left\langle \prod\nolimits_{\nu =1}^{N}\Phi _{j_{n},m_{\nu }}^{s_{\nu
}}(z_{\nu })\right\rangle =\frac{1}{n_{+}!n_{-}!}\left( \frac{\lambda }{2k}%
\right) ^{-2(j_1+...j_N)+1-N)}\prod\nolimits_{\nu =1}^{N}N_{j_{\nu },m_{\nu
}}^{s_{\nu }}\times
\end{equation*}%
\begin{equation*}
\times \prod\nolimits_{\mu <\nu }^{N}(z_{\mu }-z_{\nu })^{s_{\mu }s_{\nu
}-b^{2}(2j_{\mu }+1)(2j_{\nu }+1)}\int
\prod\nolimits_{r=1}^{n_{+}}d^{2}w_{r}\prod\nolimits_{l=1}^{n_{-}}d^{2}y_{l}%
\times
\end{equation*}%
\begin{equation*}
\times
\prod\nolimits_{l=1}^{n_{-}}\prod%
\nolimits_{r=1}^{n_{+}}(w_{r}-y_{l})^{-1-b^{2}}\prod\nolimits_{l<l^{\prime
}}^{n_{-}}(y_{l}-y_{l^{\prime }})^{1-b^{2}}\prod\nolimits_{r<r^{\prime
}}^{n_{+}}(w_{r}-w_{r^{\prime }})^{1-b^{2}}\times
\end{equation*}%
\begin{equation*}
\times \prod\nolimits_{\nu =1}^{N}\prod\nolimits_{r=1}^{n_{+}}(z_{\nu
}-w_{r})^{-s_{\nu }-b^{2}(2j_{\nu }+1)}\prod\nolimits_{\nu
=1}^{N}\prod\nolimits_{l=1}^{n_{-}}(z_{\nu }-y_{l})^{s_{\nu }-b^{2}(2j_{\nu
}+1)}\times
\end{equation*}%
\begin{equation}
\times \left\langle \prod\nolimits_{\nu =1}^{N}\gamma _{(z_{\nu })}^{m_{\nu
}-j_{\nu }-(s_{\nu }+1)/2}\prod\nolimits_{l=1}^{n_{-}}\beta
_{(y_{l})}\right\rangle _{\text{free}}\times h.c.  \label{INT}
\end{equation}
In addition, we may resort to projective invariance to fix three points at $%
z_{1}=0$, $z_{2}=1$, and $z_{N}=\infty $.

The correlator of the ($\beta $,$\gamma $) ghost fields in (\ref{INT}) yields
a rather complicated expression in general. Nevertheless, it simplifies
substantially in some particular cases. For instance, in the case of two
insertions it reads \cite{BB,GN}
\begin{equation*}
\left\langle \gamma _{(z_{1}=0)}^{m_{1}-j_{1}-(s_{1}+1)/2}\gamma
_{(z_{2}=1)}^{m_{2}-j_{2}-(s_{2}+1)/2}\prod\nolimits_{l=1}^{n_{-}}\beta
_{(y_{l})}\right\rangle _{\text{free}}\times
h.c.=\prod\nolimits_{l=1}^{n_{-}}|y_{l}|^{-2}|1-y_{l}|^{-2}\times
\end{equation*}%
\begin{equation*}
\times (-1)^{n_{-}}\frac{\Gamma (1/2-j_{1}-s_{1}/2+m_{1})}{\Gamma (1/2+j_{1}+s_{1}/2-m_{1})}\frac{\Gamma
(1/2-j_{2}-s_{2}/2+m_{2})}{\Gamma
(1/2+j_{2}+s_{2}/2-m_{2})}.
\end{equation*}%
World-sheet integral (\ref{INT}) can in principle be
computed by using generalized Selberg integral formulas of the type worked
out in \cite{DF,FH,FH2}. To do this one has to give a precise prescription for
contour integration. We will not address the details of such prescription
in this appendix.

Integral representation (\ref{B2}) gathers the residues associated to the $N$
-point correlation functions of the OSP(1$|$2) WZNW model, and after
analytic continuation in $n_{\pm }$ and $j_{i}$ the exact form of the
correlation functions are obtained. The exact expressions for two- and
three-point correlation functions in the OSP(1$|$2) WZNW model were found in
\cite{HS2}.

Representation (\ref{B2}) gives
important information about the correlators. For instance, the KPZ scaling
properties can be read from this expression. Correlators (\ref{INT}) scale
as $\sim \left( \lambda /2k\right) ^{n_{+}+n_{-}}$, where, according to (\ref%
{Dos2}), $n_{+}+n_{-}=1-2(j_{1}+j_{2}+...j_{N})-N$. In particular, for the
two-point function, where $N=2$ and $j_{1}=j_{2}=j$, we obtain $\sim \left(
\lambda /2k\right) ^{-4j-1}$. So, let us compare this with the scaling
properties of the exact exact solution of the OPS(1$|$2) WZNW model found in
\cite{HS2}. First, let us notice that in comparing the conventions of \cite%
{HS2} with ours here we have to redefine $j$ as follows $j\rightarrow j+1/2$%
. Thus, the KPZ scaling is $\sim \left( \lambda /2k\right) ^{-4j-3}$ which
precisely agrees with the result in \cite{HS2}. Actually, it is not hard to
see that if one introduces the scale $\lambda $ in the formulas of \cite{HS2}
then eq. (4.7) therein scales like
\begin{equation}
\sim \left( \frac{2kb^{2}}{i\lambda \gamma (\frac{b^{2}+1}{2})}\right)
^{4j+3}.
\end{equation}%
Analogously, for the three-point functions one finds $\sim \left( \lambda
/2k\right) ^{-2(j_{1}+j_{2}+j_{3})-5}$ which also coincides with the scaling
of eq. (4.21) of \cite{HS2}.

One can also see that in the coincidence limit $\lim_{z_{1}\rightarrow
z_{2}}\Phi _{j_{1}m\,_{1}}^{s_{1}}(z_{1})\Phi _{j_{2}m\,_{2}}^{s_{2}}(z_{2})$%
, where two of the vertices hit each other, the pole condition that appears
at $z_{1}=z_{2}$ can be interpreted as a mass-shell condition $L_0 -1 =
-2b^{2}j(j+1/2)+s(s-1)/2+l=0$ of a level-$l$ excited intermediate state carrying
momenta $j=j_{1}+j_{2}-1+(N+n_{+}^{\prime }+n_{-}^{\prime })/2$ and $%
s=s_{1}+s_{2}+n_{+}^{\prime }-n_{-}^{\prime }$, where $n_{\pm }^{\prime
}\leq n_{\pm }$ are the amount of screening operators of the type $S_{\pm }$
whose inserting points also tend to $z_{2}$ together with $z_{1}$. In this
limit, the $N$-point function factorizes in the product of one three-point
function times one $N-2$-point function.



\baselineskip=11pt

\begin{thebibliography}{99}


\bibitem{Ma}
  J.~M.~Maldacena,
  ``The large $N$ limit of superconformal field theories and supergravity,''
  Adv.\ Theor.\ Math.\ Phys.\  {\bf 2} (1998) 231
  [Int.\ J.\ Theor.\ Phys.\  {\bf 38} (1999) 1113]
  [arXiv:hep-th/9711200];


\bibitem{MT}
  R.~R.~Metsaev and A.~A.~Tseytlin,
  ``Type IIB superstring action in $AdS_5 \times S^5$ background,''
  Nucl.\ Phys.\  B {\bf 533}, 109 (1998)
  [arXiv:hep-th/9805028].

\bibitem{BVW}
  N.~Berkovits, C.~Vafa and E.~Witten,
  ``Conformal field theory of AdS background with Ramond-Ramond flux,''
  JHEP {\bf 9903}, 018 (1999)
  [arXiv:hep-th/9902098].


\bibitem{TT}
 T.~Takayanagi and N.~Toumbas,
  ``A matrix model dual of type 0B string theory in two dimensions,''
  JHEP {\bf 0307} (2003) 064
  [arXiv:hep-th/0307083].



\bibitem{DKKMMS}
  M.~R.~Douglas, I.~R.~Klebanov, D.~Kutasov, J.~M.~Maldacena, E.~J.~Martinec and N.~Seiberg,
  ``A new hat for the $c = 1$ matrix model,''
  arXiv:hep-th/0307195.


\bibitem{Ta}
  T.~Takayanagi,
  ``Matrix model and time-like linear dilaton matter,''
  JHEP {\bf 0412} (2004) 071
  [arXiv:hep-th/0411019].


\bibitem{Takayanagi}
  T.~Takayanagi,
  ``$c < 1$ string from two dimensional black holes,''
  JHEP {\bf 0507}, 050 (2005)
  [arXiv:hep-th/0503237].


\bibitem{OT}
  T.~Okuda and T.~Takayanagi,
  ``Ghost D-branes,''
  JHEP {\bf 0603} (2006) 062
  [arXiv:hep-th/0601024].

\bibitem{Berkovits1}
  N.~Berkovits,
  ``A new limit of the $AdS_5 \times S^5$ sigma model,''
  JHEP {\bf 0708}, 011 (2007)
  [arXiv:hep-th/0703282].
 
\bibitem{Berkovits2}
  N.~Berkovits and C.~Vafa,
  ``Towards a worldsheet derivation of the Maldacena conjecture,''
  JHEP {\bf 0803}, 031 (2008)
  [AIP Conf.\ Proc.\  {\bf 1031}, 21 (2008)]
  [arXiv:0711.1799 [hep-th]].
 
\bibitem{Bonelli1}
  G.~Bonelli and H.~Safaai,
  ``On gauge/string correspondence and mirror symmetry,''
  JHEP {\bf 0806}, 050 (2008)
  [arXiv:0804.2629 [hep-th]].


\bibitem{Berkovits3}
  N.~Berkovits,
  ``Perturbative super-Yang-Mills from the topological 
    $AdS_5 \times S^5$ sigma model,''
  JHEP {\bf 0809}, 088 (2008)
  [arXiv:0806.1960 [hep-th]].

\bibitem{Bonelli2}
  G.~Bonelli, P.~A.~Grassi and H.~Safaai,
  ``Exploring pure spinor string theory on 
   $AdS_4\times \mathbb{CP}^3$,''
  JHEP {\bf 0810}, 085 (2008)
  [arXiv:0808.1051 [hep-th]].


\bibitem{MV}
  S.~Mukhi and C.~Vafa,
  ``Two-dimensional black hole as a topological coset model of $c = 1$ string
  theory,''
  Nucl.\ Phys.\  B {\bf 407}, 667 (1993)
  [arXiv:hep-th/9301083].

\bibitem{BO}
  M.~Bershadsky and H.~Ooguri,
  ``Hidden OSP($N|$2) symmetries in superconformal field theories,''
  Phys.\ Lett.\  B {\bf 229}, 374 (1989).

\bibitem{BO2}
  M. Bershadsky and H. Ooguri,
  ``Hidden SL$(n)$ symmetry in conformal field theories,''
  Comm. Math. Phys. \textbf{126} (1989) 49.

\bibitem{KS}
  Y.~Kazama and H.~Suzuki,
  ``New ${\cal N}=2$ superconformal field theories and superstring compactification,''
  Nucl.\ Phys.\  B {\bf 321}, 232 (1989).

\bibitem{KS2}
  Y.~Kazama and H.~Suzuki,
  ``Characterization of ${\cal N}=2$ superconformal models generated by coset space
  method,''
  Phys.\ Lett.\  B {\bf 216}, 112 (1989).

\bibitem{CRS}
  T.~Creutzig, P.~B.~Ronne and V.~Schomerus,
  ``${\cal N}=2$ superconformal symmetry in super coset models,''
  arXiv:0907.3902 [hep-th].

\bibitem{Creutzig}
  T.~Creutzig,
  ``Branes in supergroups,''
  DESY-THESIS-2009-018.

\bibitem{HS2}
  Y.~Hikida and V.~Schomerus,
  ``Structure constants of the OSP(1$|$2) WZNW model,''
  JHEP {\bf 0712}, 100 (2007)
  [arXiv:0711.0338 [hep-th]].

\bibitem{RT}
  S.~Ribault and J.~Teschner,
  ``$H_3^+$ WZNW correlators from Liouville theory,''
  JHEP {\bf 0506}, 014 (2005)
  [arXiv:hep-th/0502048].

\bibitem{HS3}
  Y.~Hikida and V.~Schomerus,
  ``$H^+_3$ WZNW model from Liouville field theory,''
  JHEP {\bf 0710}, 064 (2007)
  [arXiv:0706.1030 [hep-th]].

\bibitem{Wakimoto}
  M.~Wakimoto,
  ``Fock representations of the affine lie algebra $A_1(1)$,''
  Commun.\ Math.\ Phys.\  {\bf 104} (1986) 605.

\bibitem{ERS}
  I.~P.~Ennes, A.~V.~Ramallo and J.~M.~Sanchez de Santos,
  ``On the free field realization of the osp(1$|$2) current algebra,''
  Phys.\ Lett.\  B {\bf 389}, 485 (1996)
  [arXiv:hep-th/9606180].

\bibitem{ERS2}
  I.~P.~Ennes, A.~V.~Ramallo and J.~M.~Sanchez de Santos,
  ``Structure constants for the osp(1$|$2) current algebra,''
  Nucl.\ Phys.\  B {\bf 491}, 574 (1997)
  [arXiv:hep-th/9610224].

\bibitem{LeClair}
  A.~LeClair,
  ``The gl(1$|$1) super-current algebra: the role of twist and logarithmic
  fields,''
  arXiv:0710.2906 [hep-th].

\bibitem{PT}
  T.~Creutzig and P.~B.~Ronne,
  ``The GL(1$|$1)-symplectic fermion correspondence,''
  Nucl.\ Phys.\  B {\bf 815}, 95 (2009)
  [arXiv:0812.2835 [hep-th]].

\bibitem{Dotsenko}
  V.~S.~Dotsenko,
  ``The free field representation of the SU(2)
conformal field theory,''
  Nucl.\ Phys.\  B {\bf 338} (1990) 747.

\bibitem{Dotsenko2}
  V.~S.~Dotsenko,
  ``Solving the SU(2) conformal field theory with the Wakimoto free field
  representation,''
  Nucl.\ Phys.\  B {\bf 358}, 547 (1991).


\bibitem{BK}
  M.~Bershadsky and D.~Kutasov,
  ``Comment on gauged WZW theory,''
  Phys.\ Lett.\  B {\bf 266}, 345 (1991).

\bibitem{DVV}
  R.~Dijkgraaf, H.~L.~Verlinde and E.~P.~Verlinde,
  ``String propagation in a black hole geometry,''
  Nucl.\ Phys.\  B {\bf 371}, 269 (1992).

\bibitem{MO}
  J.~M.~Maldacena and H.~Ooguri,
  ``Strings in $AdS_3$ and SL(2,${\mathbb R}$) WZW model. I,''
  J.\ Math.\ Phys.\  {\bf 42}, 2929 (2001)
  [arXiv:hep-th/0001053].

\bibitem{EY}
  T.~Eguchi and S.~K.~Yang,
  ``${\cal N}=2$ superconformal models as topological field theories,''
  Mod.\ Phys.\ Lett.\  A {\bf 5} (1990) 1693.

\bibitem{WTST}
  E.~Witten,
  ``Mirror manifolds and topological field theory,''
  arXiv:hep-th/9112056.

\bibitem{NN}
  S.~Nakamura and V.~Niarchos,
  ``Notes on the S-matrix of bosonic and topological non-critical strings,''
  JHEP {\bf 0510}, 025 (2005)
  [arXiv:hep-th/0507252].

\bibitem{Ribault}
  S.~Ribault,
  ``Knizhnik-Zamolodchikov equations and spectral flow in $AdS_3$ string
  theory,''
  JHEP {\bf 0509}, 045 (2005)
  [arXiv:hep-th/0507114].


\bibitem{Giribet2006}
  G.~Giribet,
  ``The string theory on $AdS_3$ as a marginal deformation of a linear  dilaton
  background,''
  Nucl.\ Phys.\  B {\bf 737}, 209 (2006)
  [arXiv:hep-th/0511252].

\bibitem{HS}
  Y.~Hikida and V.~Schomerus,
  ``The FZZ-duality conjecture - A proof,''
  arXiv:0805.3931 [hep-th].

\bibitem{DVV2}
  R.~Dijkgraaf, H.~L.~Verlinde and E.~P.~Verlinde,
  ``Topological strings in $d < 1$,''
  Nucl.\ Phys.\  B {\bf 352}, 59 (1991).

\bibitem{BCOV}
  M.~Bershadsky, S.~Cecotti, H.~Ooguri and C.~Vafa,
  ``Kodaira-Spencer theory of gravity and exact results for quantum string
  amplitudes,''
  Commun.\ Math.\ Phys.\  {\bf 165} (1994) 311
  [arXiv:hep-th/9309140].


\bibitem{FZZunpublished}
  V. Fateev, A.B. Zamolodchikov and Al. Zamolodchikov, unpublished note.

\bibitem{HoKa}
  K.~Hori and A.~Kapustin,
  ``Duality of the fermionic 2d black hole and 
 ${\cal N} = 2$ Liouville theory as
  mirror symmetry,''
  JHEP {\bf 0108} (2001) 045
  [arXiv:hep-th/0104202].

\bibitem{KKK}
  V.~Kazakov, I.~K.~Kostov and D.~Kutasov,
  ``A matrix model for the two-dimensional black hole,''
  Nucl.\ Phys.\  B {\bf 622}, 141 (2002)
  [arXiv:hep-th/0101011].

\bibitem{KPZ}
  V.~G.~Knizhnik, A.~M.~Polyakov and A.~B.~Zamolodchikov,
  ``Fractal structure of 2d-quantum gravity,''
  Mod.\ Phys.\ Lett.\  A {\bf 3}, 819 (1988).

\bibitem{BB}
  K.~Becker and M.~Becker,
  ``Interactions in the SL(2,$\mathbb{R}$) / U(1) black hole background,''
  Nucl.\ Phys.\  B {\bf 418}, 206 (1994)
  [arXiv:hep-th/9310046].

\bibitem{GN}
  G.~Giribet and C.~A.~Nunez,
  ``Correlators in $AdS_3$ string theory,''
  JHEP {\bf 0106}, 010 (2001)
  [arXiv:hep-th/0105200].

\bibitem{DF}
  V.~S.~Dotsenko and V.~A.~Fateev,
  ``Four point correlation functions and the operator algebra in the
  two-dimensional conformal invariant theories with the central charge $c < 1$,''
  Nucl.\ Phys.\  B {\bf 251}, 691 (1985).

\bibitem{FH}
  T.~Fukuda and K.~Hosomichi,
  ``Three-point functions in sine-Liouville theory,''
  JHEP {\bf 0109}, 003 (2001)
  [arXiv:hep-th/0105217].

\bibitem{FH2}
  T.~Fukuda and K.~Hosomichi,
  ``Super Liouville theory with boundary,''
  Nucl.\ Phys.\  B {\bf 635}, 215 (2002)
  [arXiv:hep-th/0202032].



\end{thebibliography}
\end{document}